\documentstyle[preprint,aps,tighten,epsf,floats]{revtex}
\newcommand{\ben}{\begin{displaymath}}
\newcommand{\een}{\end{displaymath}}
\newcommand{\be}{\begin{equation}}
\newcommand{\ee}{\end{equation}}
\newcommand{\bea}{\begin{eqnarray}}
\newcommand{\eea}{\end{eqnarray}}

\newcommand{\fign}[1]{\label{#1}}

\begin{document}
\draft
\title{Are Pions Perturbative in Effective Field Theory?}
\author{ J. Gegelia}
\address{Department of Physics, The Flinders University of South Australia, \\
Bedford Park, SA 5042, Australia}
\date{\today}
\maketitle 

\begin{abstract}
It is shown that pions can be included perturbatively into effective field
theory only for
the external momenta, well below the pion mass. But for such low energies it is
not necessary to include pions explicitly.
\end{abstract}

\pacs{
03.65.Nk,  
%Nonrelativistic scattering theory
11.10.Gh,  
%Renormalization
12.39.Fe,  
%Chiral Lagrangians
13.75.Cs.} 
%Nucleon-nucleon interactions (including antinucleons, deuterons,
%          etc.)

\section{Introduction}

Weinberg's original ideas about the chiral perturbation theory approach to
processes involving an arbitrary number of nucleons \cite{weinberg1,weinberg2}
were followed by intensive investigations of various aspects of this approach
(see for example \cite{savage} and references included therein).

For processes involving more than one nucleon Weinberg suggested to apply the
power counting to the potential rather than to the scattering amplitude. For
N-nucleon processes the
potential is defined as a sum of N-nucleon irreducible time-ordered perturbation
theory diagrams. 
The amplitude is to be found by solving Lippmann-Schwinger equation (or
Schr\" odinger equation).

% Iteration of the potential via
% the Lippmann-Schwinger equation leads to divergences. One could try to
% regularize the potential and include counter-terms, but due to the
% non-renormalizability of the
%theory one would have to include an infinite number of (counter-)terms with
%more and more derivatives. 
%So, one has either to exactly solve the equation and after subtract
%divergences explicitly, or otherwise one should draw all relevant diagrams,
%subtract them and then sum these renormalised diagrams. Unfortunately there
%is no equation for the renormalised amplitude. 
%The low order calculations 
%require inclusion of the contributions of an infinite number of counter-terms
%with
%higher and higher order derivatives (up to infinity). But this does not
%mean that Weinberg's power counting is not consistent, as was claimed in
%ref. \cite{kaplan3,savage}. Note that power counting
%should be applied to renormalised
%diagrams only, i.e. after inclusion of the contributions of counter-terms, so
%the 
%involvement of higher order counter-terms into low-order calculations does not
%affect power counting at all. More details about Weinberg's power counting are
%included in the next section. 

%Divergences appearing in diagrams can be
%regulated using (sharp or smooth) cut-off regularization. One can keep cutoff
%parameters of order of the mass of the lightest particle which was integrated
%out and fit coupling constants to
%the experimental data as was done in ref. \cite{ordonez}. 
 
Recently it was 
suggested that pions can be included perturbatively into effective field theory
calculations for momenta up to the pionic mass and even higher (up to $300 \
{\rm MeV}$ in the centre of mass)
\cite{kaplan2,kaplan3,kapland,savage,kaplan4}. In these papers dimensional
regularization and the Power Divergent Subtraction (PDS) scheme were used to
describe the
$NN$ scattering data and electro-magnetic form factors of deuteron. In the PDS
scheme
the coefficients (coupling constants) of leading and sub-leading order terms
in the effective Lagrangian for ${ }^1S_0$ wave are of the order
$\sim (100 \ {\rm MeV})^{-2}$ and $\sim (150 \ {\rm MeV})^{-4}$ respectively
for the
normalisation point equal to pion mass \cite{kaplan2}.
These values of couplings are not encouraging at all. After inclusion of the
pion 
explicitly one would expect that the scale of couplings would be determined by
the mass of the lightest integrated particle $\sim (800 \ {\rm MeV})$ provided
that the
normalisation point is taken to be of the order of the pion mass.
Impressive numerical fits given in above mentioned papers extend up to
the values of momenta where ``small'' expansion parameters are even larger than
1. All powers are equally important for such values of expansion parameter
and it raises a question whether these fits have anything to do with suggested
power counting. Of course there could be some magnificent cancellations among
higher order terms but power counting does not take into account such accidental
cancellations.  

In Weinberg's power counting the one pion exchange potential is of leading order
and 
hence it has to be iterated via the Lippmann-Schwinger equation. Below, to
investigate a little further the power counting arguments, pions are included
perturbatively using the subtractions at $p^2=-\mu^2$.
The conclusion is that using the subtractions at $p^2=-\mu^2$ pions can be
included perturbatively only for momenta
well below the mass of the pion.
Close investigations demonstrate that the same conclusions are valid for PDS
scheme too in contrast to the results of papers
\cite{kaplan2,kaplan3,kapland,savage,kaplan4}.
 
\medskip
\medskip
\medskip

\section{Explicit Calculations }

Concentrating on the $N\!N$ scattering problem one could summarise Weinberg's
ideas in the following way:

%The effective chiral Lagrangian is 
%non-renormalizable in the traditional sense but it contains all possible terms
%which are not suppressed by the symmetries of the theory and the ultraviolet
%divergences  are absorbed into the parameters of the Lagrangian.
%Renormalization points should be chosen of the order of external
%momenta $p$.
%After renormalization, the effective cut-off is of order $p$  \cite{weinberg2}.
%For processes involving more than one nucleon, one finds that at any order
%there are an infinite number of diagrams. 
%For $NN$ scattering the low order calculations 
%require inclusion of the contributions of an infinite number of counter-terms
%with
%higher and higher order derivatives (up to infinity). But this does not
%mean that Weinberg's power counting is not consistent, as was claimed in
%ref. \cite{kaplan3,savage}. Note that power counting
%should be applied to renormalized
%diagrams only, i.e. after inclusion of the contributions of counter-terms, so
%the 
%involvement of higher order counter-terms into low-order calculations does not
%affect power counting at all.  

One should draw all the diagrams for a given process (there will be an infinite
number of them). Diagrams with loops will contain divergences.  The effective
chiral Lagrangian \cite{ordonez} is non-renormalizable in the traditional sense
where only a finite number of parameters are involved; rather, it contains all
possible terms which are not suppressed by the symmetries of the theory with the
ultraviolet divergences being absorbed into an infinite number of parameters of
the Lagrangian.  One removes divergences by subtracting diagrams (or
equivalently one could include contributions of counter-terms) and taking
normalisation points of the order of external momenta.  For these subtracted
diagrams the effective cut-off is then of the order of external momenta.  Once
these diagrams have been renormalised, one can sort them by orders of the small
expansion parameter (external momentum, renormalization point, or mass of pion)
by applying Weinberg's power counting rules.

It turns out that the renormalised diagrams contributing up to and including any
given order $n$ in the small parameters consist of a finite number of diagrams
which are two-nucleon irreducible, and an infinite number of diagrams which are
two-nucleon reducible. Denoting the sum of the contributing irreducible diagrams
by $V^R$ and the sum of the same diagrams before renormalization formally by
$V$, one finds that the series
\be T^R = V^R + (VG_0V)^R + (VG_0VG_0V)^R + \ldots
\ee
includes all contributions up to order $n$ (and some of the contributions of
order greater than $n$). Here $G_0$ is the free two-nucleon Green function and
each iteration of $V$ is renormalised separately (as indicated by a superscript
$R$).  The summation of this series is highly non-trivial, e.g., there is no
known
way of writing down a closed integral equation for $T^R$. To make progress
one can pursue one of the following:
\begin{enumerate}
\item

One could try to solve the Lippmann-Schwinger (LS) equation $T=V+VG_0T$
analytically with a formal un-subtracted non-regularized potential, or more
rigorously, one could regularize the loop integrals in $V$ and $VG_0T$ and then
solve the LS equation analytically. In both cases one would need to perform
subtractions (renormalization) in the resulting solutions. 

Although analytic solutions to the LS equations are rare, such approaches have
been successfully applied to the $ { }^1S_0$ NN problem with (only) contact
interaction terms \cite{weinberg2,gegelia}. For this simple case it is easy to
verify the general assumption that solving the LS equation with regularization
and after subtracting divergences in the solution is equivalent to the summation
of already subtracted diagrams.

Solving the LS equation and after subtracting divergences is equivalent to
summing an infinite number of renormalised diagrams. Note that diagrams of all
orders (up to infinity) are involved in this summation. The higher order
subtracted diagrams include contributions of higher order counter-terms. For
example, if one takes the sum of the first two (leading and sub-leading) terms
of 
the potential in the chiral expansion and iterates it using the LS equation, and
then renormalises the obtained amplitude, one thereby includes the contributions
of an infinite number of counter-terms with all orders of derivatives (up to
infinity).  As far as power counting must be performed for renormalised
diagrams, i.e. after taking into account the contributions of counter-terms, the
presence of contributions of an infinite number of counter-terms does not change
power counting arguments despite the claims of ref.\cite{luke,kaplan3,savage}
that Weinberg's power counting is inconsistent. A similar situation arises in
the meson sector. As mentioned in Ref.\ \cite{gasser}, high order terms (high
order loops) contribute at low order for calculations (before renormalization)
of $\pi$-$\pi$ scattering, if one does not make use of dimensional
regularization. Does this mean that power counting is not valid in other
regularizations and that different regularizations are not equivalent? Of course
not. Power counting is valid {\em after} renormalization, i.e. for subtracted
diagrams, and hence for regularizations different from dimensional
regularization one just has to perform some additional subtractions (over and
above what is needed when dimensional regularization is used). One gets the same
renormalised diagrams with the same power counting provided that the
normalisation conditions are the same.

\item
One could utilise cut-off theory \cite{ordonez,park,gegelia2,lepage,lepage1}.
The advantage of cut-off theory is that one can solve the LS equation
numerically. It is interesting to note that the aforementioned criticism
\cite{luke,kaplan3,savage} of Weinberg's power counting has recently been shown
not to apply to the cut-off theory \cite{park}. The discussion given in Ref.\
\cite{park} and the one above for renormalised theory correspond with each
other, supporting the equivalence of renormalised and cut-off theory once again
\cite{gegelia2}.

\item
And of course one could try to sum renormalised diagrams directly.
\end{enumerate}

To follow the third way one can use the EFT expansion of the quantity
$pcot\delta({\bf p})$
suggested in \cite{kaplan}:
\begin{equation}
pcot\delta({\bf p})=ip+{4\pi\over M}{1\over {\cal A}}=ip+{4\pi\over M}
\left( {1\over
{\cal A}_0}-{{\cal A}_2\over {\cal A}_0^2}+...\right) 
\end{equation}
where ${\cal A}_0$ is non-perturbative amplitude, which is a sum of all diagrams
with leading order potential as a vertex. ${\cal A}_2$ contains all
contributions of leading order potential with one insertion of second order
potential etc. \cite{kaplan}. One could solve Lippmann-Schwinger equation for
the potential with leading and second order contributions and find the inverse
amplitude. The difference with ${1\over
{\cal A}_0}-{{\cal A}_2\over {\cal A}_0^2}$ would be of higher order and hence
the agreement between this two solutions should be good up to the validity of
considered 
approximation. The large difference would be an indicator that diagrams with
many
insertions of second order potential are larger than estimated. One can include
part of
second (or higher) order potential into ${\cal A}_0$ non-perturbatively, while
including another part as perturbative insertions. As was mentioned in the
introduction,
power counting arguments do not rely on cancellation between diagrams of the
same order, so the higher order contributions coming from that part of second
order potential which was included non-perturbatively, should be small.

The contribution of contact interaction terms to
the 2-nucleon potential in the partial ${ }^1S_0$ wave up to (and including)
the sub-leading order has the form:
\begin{equation}
V^{(2)}_C\left( {\bf p},{{\bf p}'}\right)=V_0+V_2=C_0+C_2\left({\bf p}^2+
{\bf p'}^2\right)
\end{equation}

The one pion exchange potential in the ${ }^1S_0$ channel is:
\begin{equation}
V_{\pi}\left( {\bf p},{\bf p}'\right)=-{g_A^2\over 2f^2} \  {{\bf q}^2\over
{\bf q}^2+m_\pi^2}=-{g_A^2\over 2f^2}\left( {m_\pi^2\over
{\bf q}^2+m_\pi^2}-1\right)
\label{opep}
\end{equation}
where ${\bf q}={\bf p}'-{\bf p}$, $g_A=1.25$ and $f=132$ MeV.

According Weinberg's power counting criteria this potential is of leading order 
for the momenta of the order of the mass of pion and it is of the
order $p^2/m_\pi^2$ for the momenta much below the pion mass, provided that
the normalisation point is taken of the order of external momenta. 

\begin{figure}[t]
\hspace*{2cm}  \epsfxsize=12cm\epsfbox{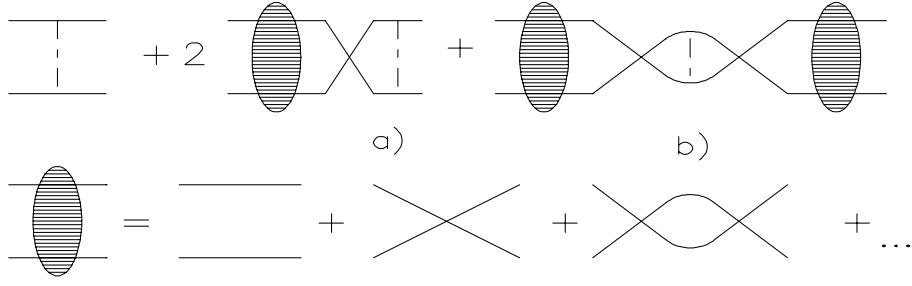}
\vspace{2mm}
\caption{\fign{feyn}{\it Graphs corresponding to ${\cal A}_1$. Dashed line
is the exchange of pion and the four solid line vertex corresponds
to the potential $V_C$.}}
\end{figure}

One can include second order contact interaction potential non-perturbatively,
pionic potential perturbatively and determine the range of validity of
perturbative inclusion of pionic potential.  
Diagrams including contact interaction and one pion are drawn in
FIG.1. The diagrams in the second row correspond to pure contact
interactions. These bubble-chain diagrams are divergent and are 
to be subtracted. They can be summed up by iterating the contact interaction
potential $V_C^{(2)}$ and subtracting at $p^2=-\mu^2$ \cite{gegelia}. 

One gets
the following expression for the non-perturbative Feynmann amplitude:
\begin{equation}
{\cal A}_0^{(2)}=-{C_0+C_2p^2\over 1+\left( C_0+C_2p^2\right){M(\mu +
ip)\over 4\pi}}
\label{a0}
\end{equation} 
The sub-diagram of diagram a), containing the pionic line, is not divergent but
has to be subtracted. The point is that the corresponding diagram in
relativistic theory is
divergent and has to be subtracted. The subtraction of the
relativistic diagram at the point ${\bf p}^2=-\mu^2$ automatically leads to the
subtraction of its non-relativistic approximation. A bubble sub-diagram of the
diagram b) containing the pionic line has two non-divergent sub-diagrams itself.
According to the
same argument as above these sub-diagrams, although finite, have to be
subtracted
before the divergent bubble sub-diagram is subtracted. Note that these
subtractions of non-divergent diagrams are necessary not to violate unitarity.

\begin{figure}[t]
\hspace*{0.2cm}  \epsfxsize=16cm\epsfbox{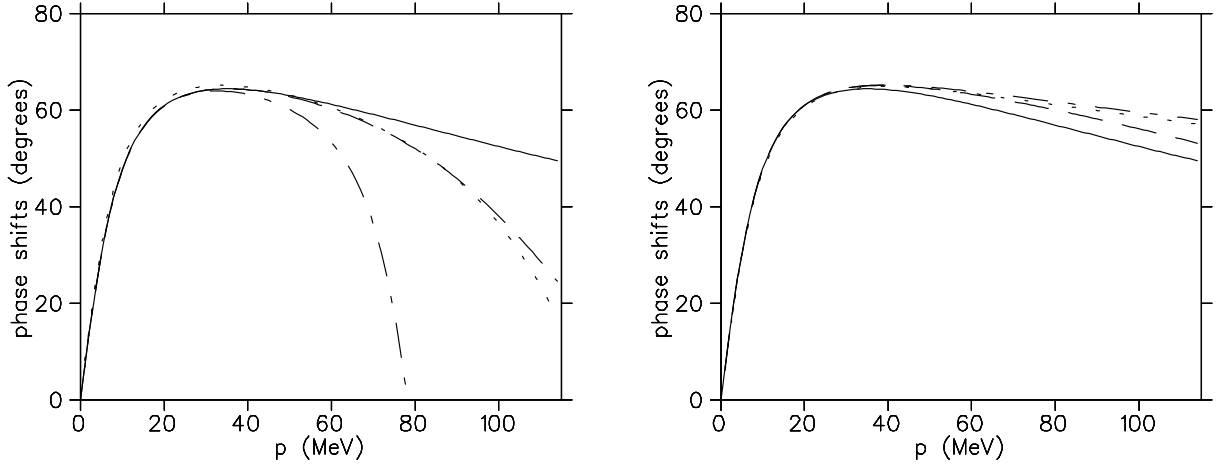}
\vspace{2mm}
\caption{\fign{cutshedareba} {\it  Phase shifts for the subtractions at
$p^2=-\mu^2$. Left figure corresponds to the non-perturbative and right figure
to the perturbative inclusion of $C_2\left({\bf p}^2+
{\bf p'}^2\right)$ term. Solid line corresponds to the effective range
expansion. Dash-dotted and dotted lines correspond to $\mu =10$ and $40$
{\rm MeV}
respectively and dashed line
corresponds to the floating normalisation point.}}
\end{figure}

After subtractions one gets the following expression:
\begin{equation}
{\cal A}_1={\cal A}_1^{(I)}+{\cal A}_1^{(II)}+{\cal A}_1^{(III)}
\label{a2}
\end{equation}
where                                                                
\begin{eqnarray}\label{17}
 {\cal A}_1^{(I)} &=&  \left({g_A^2\over 2f^2}\right)\left[ -1+{m_\pi^2\over
4p^2} \ln \left( 1 + {4p^2\over m_\pi^2}\right)\right]\ ,\nonumber\\
 {\cal A}_{1}^{(II)} &=& {g_A^2\over f^2} \left( {m_\pi M{\cal A}_0\over
4\pi}
\right)\left\{ -{(\mu +ip)\over m_{\pi}}+{m_\pi\over 2p} \left[\tan^{-1} 
\left({2p\over m_\pi}\right) +
{i\over 2} \ln
\left(1+ {4p^2\over m_\pi^2} \right)\right] -{m_\pi\over 2\mu}\ln 
\left( 1+{2\mu\over m_\pi}\right)\right\}\ ,\nonumber\\
{\cal A}_1^{(III)} &=& {g_A^2\over 2f^2} \left({m_\pi M{\cal A}_0\over
4\pi}\right)^2\times  \nonumber\\
 &{ }& \left\{  -\left( {\mu +ip\over m_{\pi}}\right)^2+
\left[ i\tan^{-1} \left({2p\over m_\pi}\right) -{ip\over \mu}
\ln \left( 1+{2\mu\over m_\pi}\right)- {1\over 2} \ln
\left( 1+ {4p^2\over m_{\pi}^2}\right)\right]\right\} 
\end{eqnarray}
and
$$
{\cal A}^{-1}=-{1\over C_0+C_2p^2}-{M\over 4\pi}\left( \mu +ip\right)-
{g_A^2\over f^2}\left[ 
{m_\pi^2\over 8p^2}\ln\left( 1+{4p^2\over m_\pi^2}\right)\left( {1\over C_0}+
{M\mu\over 4\pi}\right)^2-{1\over 2C_0^2}\right]
$$
\begin{equation}
+{g_A^2\over f^2}\left[{m_\pi^2M^2\over 128\pi^2}\ln\left( 1+
{4p^2\over m_\pi^2}\right)+{m_\pi^2M\over 4\pi}\left( {1\over C_0}+
{M\mu\over 4\pi}\right)\left( {1\over 2p}\ \tan^{-1}\left( 
{2p\over m_\pi}\right)-
{1\over 2\mu}\ \tan^{-1}\left( {2\mu\over m_\pi}\right)\right)\right]
\label{A_1}
\end{equation} 
While the whole amplitude ${\cal A}$ does not depend on $\mu $, the approximate
expression (\ref{A_1}) does and by good choice of the value of this parameter
one should make contributions of higher order terms small.
Matching to the effective range expansion one can determine $C_0$ and $C_2$ for
particular values of normalisation point $\mu$. Using these values for $C_0$ and
$C_2$, one can find the phase shifts from (\ref{A_1}). The phase shifts for
$\mu =10$ MeV and $\mu =40$ MeV are plotted in FIG.2. One could choose the
most 
natural ``floating'' normalisation condition $\mu\sim p$. The phase shifts for
$\mu =p$ 
are also plotted in FIG.2. It is seen that for $\mu =40$ MeV and for the
floating
normalisation condition the agreement between the effective range expansion
and EFT
is quite satisfactory for momenta up to 60 MeV while the deviation is
significant for higher momenta. The values of couplings for $\mu =40$ MeV
$C_0\sim (57 \ {\rm MeV})^{-2}$, $C_2\sim (80 \ {\rm MeV})^{-4}$ are almost
satisfactory.
For
$\mu \sim 140 \ {\rm MeV}$ one expects couplings to be much smaller,
$\sim (800 {\rm MeV})^n$ ($n$ is determined by the dimension of coupling
constant).
Note that for such
$\mu $ pions can not be included perturbatively. 
The fit to the effective range expansion is
also unsatisfactory for a normalisation point exceeding 60 MeV. 

One can include second order contact interaction potential perturbatively and
calculate phase shifts. The results are plotted in FIG.2. It is seen that the
difference between phase shifts for perturbatively and non-perturbatively
included second order contact interaction potential is significant for momenta
above $\sim 50 {\rm MeV}$ indicating that mentioned potential should be included
non-perturbatively.  

\begin{figure}[t]
\hspace*{5cm}  \epsfxsize=8cm\epsfbox{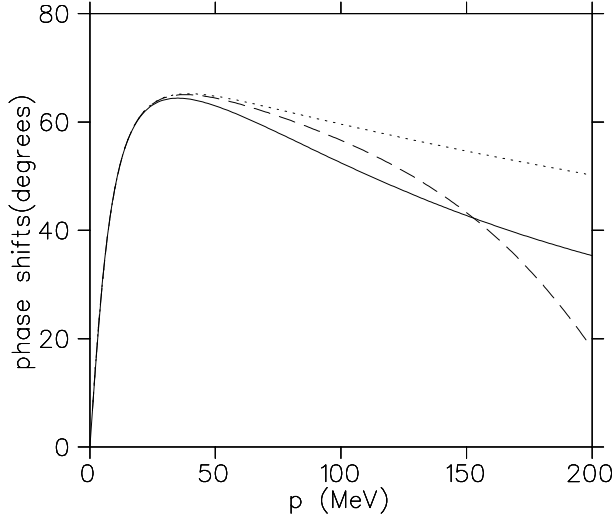}
\vspace{2mm}
\caption{\fign{mserefl}{\it Phase shifts in PDS scheme. Solid line
corresponds to the effective range expansion. 
Dashed and dotted lines correspond to non-perturbative and perturbative
inclusion of $C_2\left( p^2+p'^2\right)$ term respectively.}}
\end{figure}

\begin{figure}[t]
\hspace*{5cm}  \epsfxsize=8cm\epsfbox{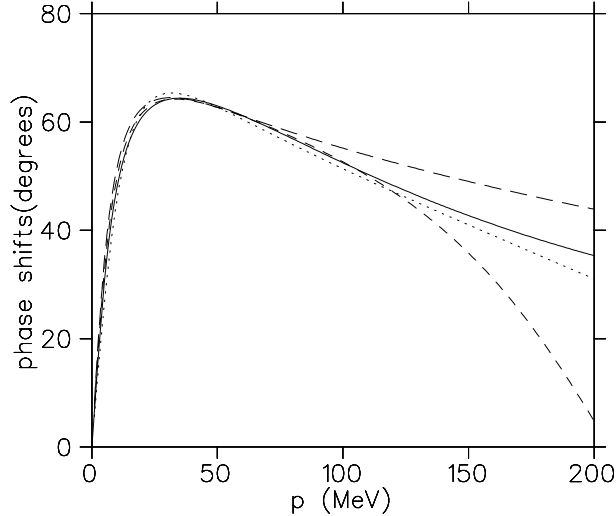}
\vspace{2mm}
\caption{\fign{mserefl}{\it 
Phase shifts in PDS scheme for ``best fit'' parameters. Solid line
corresponds to the effective range expansion. 
Dotted line corresponds to the ``best fit'' by KSW, long-dashed and dashed
lines correspond
to perturbative and non-perturbative inclusion of second order small contact
terms ($Dm_{\pi}^2$ and $C_2\left
( p^2+p'^2\right)$ terms) respectively.}}
\end{figure}

The above results are not surprising. One could  hardly expect perturbative
inclusion of pions to be satisfactory unless ${p^2/m_{\pi}^2}$ is
sufficiently small. But these conclusions do not agree with the results of ref.
\cite{kaplan2,kaplan3,kapland,savage,kaplan4}. Should one conclude that PDS is
a much
better scheme than the subtraction at $p^2=-\mu^2$?

In \cite{kaplan2} second order contact interaction and pionic potentials were
included perturbatively and also to make the amplitude $\mu$-independent (up to
the desired order)
the contact interaction vertex proportional to $D_2m_{\pi}^2$ was introduced and
included perturbatively. 

The inclusion of some terms perturbatively
makes sense only if their higher order contributions are small. Note once again
that power counting does not rely on cancellations between different
contributions, so the higher order contributions of low order terms should be
small themselves. It is quite easy
to include small contact terms (second order contact interaction term and the
term  proportional to $Dm_\pi^2$) non-perturbatively while pionic potentioal is
taken perturbatively. 
Doing so
one is lead to the diagrams depicted in FIG.1. It is straightforward to apply
PDS scheme to these diagrams.

Fitting parameters to the effective range expansion (normalisation point is
taken equal to $m_\pi$) one gets the phase shifts
for ${ }^1S_0$ which are plotted in FIG.3 together with the phase shifts got
from the perturbative inclusion of second order contact interaction term
($Dm_\pi^2$ included non-perturbatively) and phase shifts from effective range
expansion. It is seen that difference between phase shifts for perturbatively
and non-perturbatively included second order contact term 
is significant for
momenta $\sim 100 {\rm MeV}$, demonstrating that for such momenta the mentioned
term should be included non-perturbatively. Moreover the phase shifts got
using PDS scheme deviate from effective range expansion results already at
$\sim 30 {\rm MeV}$.  

%FIG.3 demonstrats that the inclusion of pions
%perturbatively  is possible only for very low momenta.  

It was mentioned in \cite{kaplan2} that the fit to the effective range
parameters reproduces the
data very well up to centre of mass momentum $p\sim 150 \ {\rm MeV}$. However
this observation is not quite reliable. 
In \cite{kaplan2} the phase shifts were determined using formula:
\be
\delta ={1\over 2i}\ln \left( 1+ i{Mp\over 2\pi}{\cal A}\right)
\label{log}
\ee
Expanding both sides to a given order with 
$\delta =\left( \delta_{(0)}+\delta_{(1)}+...\right) $ the quantities of the
same order were equated to each other thus determining the phase shifts up to
the  desired order.  
Such an expansion makes sense only if higher order terms (including high degrees
of low order terms which occur due to the expansion of $\ln$ function) are
small. One can
check for which values of external momenta is this condition satisfied just
comparing phase shifts obtained by expanding in powers of small parameter with
ones obtained by solving $\delta$ analytically. Large difference between these
two results is indication that higher order terms are not negligible by any
means. One could think that higher degrees of low order amplitude
(obtained while expanding $\ln$-function) which turned out to be large will be
cancelled by higher order terms of the amplitude, but this immediately means
that these higher order terms of the amplitude are large.
If such a cancellation takes place, it is just accidental and has nothing to do
with power counting.
  
Substituting numerical values with best fit from \cite{kaplan2} for coupling
constants $C_0\left( m_\pi\right)=-3.34 \ {\rm fm}^2$,
$D_2\left( m_\pi\right)=-0.42 \ {\rm fm}^4$, $C_2\left( m_\pi\right)=3.24 \ {\rm
fm}^4$ one gets the phase shifts for perturbatively included small contact
terms using
exact formula (\ref{log}) and also the above mentioned expansion of the exact
formula in powers of small parameter. The results are plotted
in FIG.4 together with the phase shifts defined from exact formula for
non-perturbatively included small contact
terms. It should be clear that the ``best fit'' obtained in above mentioned
papers is not reliable at all for momenta $\sim 100 \ {\rm MeV}$ (and higher)
and hardly best for lower momenta.

From FIG.3 and FIG.4 one sees that small contact terms become quite large
already at the momenta $\sim 60 \ {\rm MeV}$. After inclusion of pions
explicitly one expects these terms to become non-perturbative for considerably
higher momenta. The only explanation of this failure of effective field theory
approach is that pions can be included perturbatively only for very low external
momenta. FIG.3 suggests that it is still questionable whether the inclusion of
pions perturbatively within PDS scheme is consistent for any values of momenta
(good fit for extremally low momenta could be just due to the fit to low order
parameters).

The very reasonable explanation of inconsistency of perturbative inclusion of
the pion within  PDS scheme could be given by noting that
PDS scheme like $\overline{MS}$ puts the scale of the loop integrals equal
either 
to the mass of the pion or the normalisation point. So the expansion parameter
when pions are included perturbatively and the normalisation point is taken
equal to the pionic mass is
$\sim {m_\pi/\Lambda_{NN}}$ ($\Lambda_{NN}\sim 300 \ {\rm MeV}$) even for very
low external momenta, as was
mentioned in ref. \cite{kaplan3}. 
In this work it was pointed that the Yukawa piece of the one pion exchange
interaction supports a bound state (and consequently becomes non-perturbative)
only when ${m_\pi/\Lambda_{NN}}\sim 1.7$.  But the presence of
$\delta$-function part of one pion exchange interaction, which is known to
describe attraction \cite{cohenc55}, makes entire potential
completely different from Yukawa piece, so the above argument is hardly
relevant. 

The related problems encountered while including pions perturbatively within PDS
scheme were addressed in recent papers by J.V.Steele and R.J.Furnstahl
\cite{steel}
and T.D. Cohen and J.M.Hansen \cite{8038}.

\section{Conclusions}

For external momenta well below the mass of the pion one could include pions
perturbatively  in
effective field theory using subtractions at $p^2=-\mu^2$. For such low energies
effective theory
fits ${ }^1S_0$ wave $NN$ scattering data quite well but the explicit inclusion
of pions is not necessary. 
For higher energies (external momenta) it is necessary to include pionic
potential non-perturbatively (iterating Lippmann-Schwinger or Schr\" odinger
equation) in accordance with  Weinberg's power counting. 

The close examination demonstrates that the perturbative inclusion of the pion
into $NN$ problem within dimensional regularization combined with
Power Divergent Subtraction scheme is also consistent only for very low
energies.

One should conclude that the cutoff effective field theory
based on Weinberg's power counting (see for ex.
\cite{ordonez,park,gegelia2,lepage,lepage1}) still remains the
only systematic way of incorporation of Weinberg's ideas for
not very low energies.          

\medskip
\medskip
\medskip

{\bf ACKNOWLEDGEMENTS}

I would like to thank B.Blankleider 
% and N.Clisby
for commenting on the manuscript. 

This work was carried out whilst the author was a recipient of an 
 Overseas Postgraduate
Research Scholarship and a Flinders University Research Scholarship
at the Flinders University of South Australia.

%\newpage 
%\begin{thebibliography}{99} 

\end{document}